\RequirePackage{ifpdf}
\ifpdf 
\documentclass[pdftex]{sigma}
\else
\documentclass{sigma}
\fi

\numberwithin{equation}{section}

\begin{document}


\renewcommand{\PaperNumber}{091}

\FirstPageHeading

\ShortArticleName{A Method for Weight Multiplicity Computation Based on Berezin Quantization}

\ArticleName{A Method for Weight Multiplicity Computation\\ Based on Berezin Quantization}

\Author{David  BAR-MOSHE}

\AuthorNameForHeading{D.~Bar-Moshe}

\Address{Dune Medical Devices Ltd., P.O. Box 3131, Caesarea Industrial Park,  Israel}
\Email{\href{mailto:david@dunemedical.com}{david@dunemedical.com}}

\ArticleDates{Received July 26, 2009, in f\/inal form September 16, 2009;  Published online September 25, 2009}

\Abstract{Let $G$ be a compact semisimple Lie group and $T$ be a maximal torus of $G$. We describe a method for weight multiplicity computation in unitary irreducible representations of $G$, based on the theory of Berezin quantization on $G/T$.
Let $\Gamma _{\rm hol}( \mathcal{L}^{\lambda }) $ be the reproducing kernel Hilbert space of holomorphic sections of the homogeneous line bundle $\mathcal{L}^{\lambda }$ over $G/T$ associated with the highest weight $\lambda $ of the irreducible representation $\pi _{\lambda }$ of $G$. The multiplicity of a weight $ m $ in $\pi _{\lambda }$ is computed from functional analytical structure of the Berezin symbol of the projector in $\Gamma _{\rm hol}( \mathcal{L}^{\lambda }) $ onto subspace of weight $ m $. We describe a method of the construction of this symbol and the evaluation of the weight multiplicity as a rank of a~Hermitian form. The application of this method is described in a number of examples.}


\Classification{22E46; 32M05; 32M10; 53D50; 81Q70}

\medskip

\rightline{\it Dedicated to the memory of Professor Michael Marinov}

\section{Introduction}\label{intro}

Computation of weight multiplicities is a necessary step to the construction
of compact semi\-simple Lie group representations. There are several known formulas and methods for the computation of weight multiplicities, such as Freudenthal
recursive formula~\cite{O'Raifeartaigh-1}, the Kostant formula~\cite{Kostant-1}, the Littelman's path model \cite{Littelmann-1}, and the Billey--Guillemin--Rassart vector partition function \cite{BGR-1}. Also, geometric quantization and the orbit method of\/fer many elegant multiplicity formulas for group actions, see for example \cite{H-1,GS-1,V-1, V-2}. The combination of the combinatorial and geometric methods leads to algorithms for multiplicity computation \cite{BBCV-1}.

The present work is also based on quantization methods on coadjoint orbits, namely the Berezin quantization \cite{Berezin-1} on $G/T$, however, the method of multiplicity computation, presented here, does not involve a direct combinatorial computation but is rather based on the functional analytical structures of the reproducing kernels on the quantization spaces.

The application of Berezin quantization to the restriction of a unitary irreducible representation of a compact Lie group to a closed subgroup has been studied before in
\cite{ARN-1, BOU-1}. The work in the present paper has been generalized and extended in \cite{CAH-1}, which is mainly based on \cite{ARN-1, BOU-1} and on the arXiv version of the present work.

Let $\lambda $ be a dominant weight of the Lie algebra $\mathfrak{g}$ of  $G$ and $\pi _{\lambda }$ be the corresponding unitary irreducible representation of~$G$. Let $\mathcal{L}^{\lambda }$ be the associated homogeneous holomorphic line bundle over $G/T$, then according to the Borel--Weil theorem \cite{Hurt-1}, the reproducing kernel Hilbert space $\Gamma _{\rm hol}( \mathcal{L}^{\lambda }) $ of holomorphic sections of $\mathcal{L}^{\lambda }$ is a $G$-irreducible carrier space of the representation $\pi _{\lambda }$.

Berezin quantization (in its generalized version on spaces of sections of line bundles \cite{Peetre-1}) leads to the realization of $\pi _{\lambda }$  \cite{Marinov-1} in terms of the Berezin symbols \cite{Berezin-2} which act as integration kernels on $\Gamma _{\rm hol}( \mathcal{L}^{\lambda }) $.

We describe a method for the construction of the Berezin's symbol of projector in $\Gamma _{\rm hol}( \mathcal{L}^{\lambda }) $ onto the subspace of weight $ m $. The functional analytic structure of this projector expressed in the af\/f\/ine coordinates of the big Schubert cell of $G/T$ enables the computation of the weight multiplicity of $ m $ as the rank of a Hermitian form with rational coef\/f\/icients.

The main results of the present work are given in the following propositions:

\begin{proposition}\label{prop-1}
Let $G$ be compact semisimple Lie group, $T$ a maximal torus of $G$, and $G/T$ its fundamental projective space. Let $\pi _{\lambda }$ be the
irreducible unitary representation of $G$ of highest weight $\lambda $.
Then the multiplicity $\gamma ^{\lambda }( m) $ of the vector of
weight~$m$ in the representation space of~$\pi _{\lambda }$ is given by:
\begin{gather*}
\gamma ^{\lambda }( m) =\frac{N^{\lambda }}{V}\!\int_{G/T} \int_{T}
( m-\lambda ) ( h) \exp \big( K^{\lambda }( h\cdot z,
\overline{z}) \big) \exp \big({-}K^{\lambda }( z,\overline{z}%
) \big) d\mu ( h) d\mu ( z,\overline{z}),\!\!\!
\end{gather*}
where $N^{\lambda }$ is the dimension of the irreducible representation $%
\pi _{\lambda }$, $d\mu ( z,\overline{z}) $ is the Liouville
measure on $G/T$, $V$ is its total mass, $d\mu ( h) $ is the
Lebesgue measure on $T$ $($normalized to unit total mass$)$, $\mu (
h) $ is the character representation of $h\in T$ corresponding to
the integral weight $\mu $ and $K^{\lambda }( \zeta ,\overline{z}%
) :G/T\times G/T\longrightarrow \mathbb{C}$ is the analytic
continuation of a K\"ahler potential associated with the first Chern class of
the holomorphic homogeneous line bundle $\mathcal{L}^{\lambda }$ over $G/T$.
\end{proposition}

The application of this formula to weight multiplicity computation can be
simplif\/ied due to the following two propositions:

\begin{proposition}\label{prop-2}
The Berezin principal symbol of the projector onto the subspace of weight $m$, in the representation space of $\pi _{\lambda }$, is given by:
\begin{gather*}
L_{m}^{\lambda }( \zeta ,\overline{z}) =\int_{T}( m-\lambda
) ( h) \exp \big( K^{\lambda }( h\cdot \zeta ,\overline{z}%
) \big) d\mu ( h).
\end{gather*}
\end{proposition}

The Berezin principal symbol of the projector onto the subspace of weight $m$, restricted to the largest Schubert cell $\Sigma _{s}$ of $G/T$ is polynomial in the af\/f\/ine coordinates of $\Sigma _{s}$ in both of its arguments \cite{Marinov-1, Marinov-2}.  This parametrization is used to compute the multiplicity of the weight $m$ in~$\pi _{\lambda }$ as follows:

\begin{proposition}\label{prop-3}
Let $d^{\alpha }$ be the polynomial degree of coordinate $z^{\alpha }$ in
the restriction of $L_{m}^{\lambda }( \zeta ,\overline{z}) $ to
the largest Schubert cell $\Sigma _{s}$ of $G/T$. Then the monomials:
\begin{gather*}
u_{n}=\prod\limits_{\alpha \in \Delta ^{+}}( \zeta ^{\alpha })
^{n_{\alpha }}, \qquad  v_{n}=\prod\limits_{\alpha \in \Delta ^{+}}(
z^{\alpha }) ^{n_{\alpha }}, \qquad 0\leq n_{\alpha }\leq d_{\alpha }, \\
n=( n_{\alpha _{1}},n_{\alpha _{2}},\dots, n_{\alpha _{D}}) , \qquad
D=\dim_\mathbb{C} G/T ,
\end{gather*}
define a biholomorphic transformation $f:\Sigma _{s}\times \Sigma
_{s}\longrightarrow \mathbb{V\times }$ $\mathbb{V}$, $\mathbb{V\cong C}%
^{d}$,  $d=\sum_{\alpha \in \Delta ^{+}}d_{\alpha }$. $(\Delta ^{+}$~is the set of positive roots of $\mathfrak{g}.)$ Let $\mathbf{L}_{m}^{\lambda }( u,\overline{v}) $  be the unique
Hermitian form $($linear in the first argument and antilinear in the second$)$,
such that: $( f\ast \mathbf{L}_{m}^{\lambda }) ( \zeta ,%
\overline{z}) =L_{m}^{\lambda }( \zeta ,\overline{z}) $,
then multiplicity is given by:
\begin{gather*}
\gamma ^{\lambda }( m) = \mathrm{rank}_{\mathbb{V}}\big( \mathbf{L}%
_{m}^{\lambda }\big).
\end{gather*}
\end{proposition}

In Section~\ref{Ber-BW}, a brief review of the application of Berezin quantization on $G/T$ to the construction of unitary irreducible representations of $G$, is
given. In Section~\ref{WMF}, the formula of the projector onto the subspace of a given weight and the integration formula for weight multiplicity based the Berezin theory
are developed. In Section~\ref{WMcomp}, the method of computation of a~weight
multiplicity, based on Proposition~\ref{prop-3}, is described. In Appendix~\ref{Exmps}, two
examples of the application of the present method are given.

\section[Berezin quantization and the Borel-Weil theory]{Berezin quantization and the Borel--Weil theory}\label{Ber-BW}

\subsection{Notations}
\begin{enumerate}\itemsep=0pt
\item[]$G$ is a compact semisimple Lie group;

\item[]$T$ is a maximal torus of $G$;

\item[]$G^{c}$  is the complexif\/ication of $G$;

\item[]$G/T$ is a fundamental projective pace of $G$;

\item[]$W$  is the Weyl group of $G$;

\item[]$B$   is a Borel subgroup of containing $T$;

\item[]$N$ is the unipotent radical of $B$;

\item[]$N_{-}$ is the subgroup of opposite to $N$ ($N_{-}=sNs^{-1}$,where $s\in  {\rm Norm}( T) $ is any representative of the unique maximal length element of~$W$);

\item[] $\mathfrak{g}$ is the Lie algebra of $G$;

\item[] $\mathfrak{t}$ is the Cartan subalgebra of $\mathfrak{g}$ corresponding to $T$;

\item[] $\mathfrak{g}^{c}$ is the Lie algebra of $G^{c}$;

\item[] $\mathfrak{n}$ is the Lie algebra of $N$;

\item[] $\mathfrak{n}_{-}$ is the Lie algebra of $N_{-}$;

\item[] $\mathfrak{g}^{\ast }$  is the dual space of $\mathfrak{g}$;

\item[] $ \langle \cdot ,\cdot \rangle $ is the duality map $\mathfrak{g}^{\ast }\times \mathfrak{g}\longrightarrow \mathbb{C}$;

\item[] $\Delta \subset \mathfrak{g}^{\ast }$ is the set of roots of $\mathfrak{t}$   in $%
\mathfrak{g}^{c}$;

\item[] $\Delta ^{+}\subset \Delta $ is the set of positive roots (we choose $\Delta
^{+}$ as the set of roots of $\mathfrak{n}_{-}$);

\item[] $\Sigma \subset \Delta ^{+}$  is the set of primitive roots ($\Sigma
=\left\{ \gamma ^{j},j=1,\dots ,\mathrm{rank}( \mathfrak{g}) \right\}$);

\item[] $\left\{ E_{\alpha },\alpha \in \Delta ^{+}\right\} $ is the set of positive
root generators of (generators of $\mathfrak{n}_{-}$);

\item[] $\mathcal{W}$   is the Weight lattice of $G$ ($\lambda \in \mathcal{W}$
is an integral weight of $T$);

\item[] $\mathcal{C}\subset \mathcal{W}$  is the positive Weyl chamber $\mathcal{W}
$ ($\mathcal{C=}\left\{ \lambda \in \mathcal{W}  \mid  \lambda \cdot \alpha
\geq 0, \ \forall\;  \alpha \in \Delta ^{+}\right\} $,
where $\lambda \cdot \alpha $ denotes the inner product induced on $\mathcal{W}$ by the
Cartan--Killing form);

\item[] $\left\{ w_{j}\in \mathcal{W}, \ j=1,\dots ,\mathrm{rank}( \mathfrak{g}%
) \right\} $  is the set primitive weights ($w_{j}\cdot \gamma
^{i}=\delta _{j}^{i}$).
\end{enumerate}

\subsection{The quantization space in Berezin theory}

Let $\mathcal{L}^{\lambda }=( G^{c}\times \mathbb{C}^{\lambda
}) /B$ be the homogenous holomorphic line bundle over $G/T$ associated
with the dominant weight $\lambda $, def\/ined as the set of equivalence
classes of elements of $G^{c}\times \mathbb{C}$ under the equivalence
relation:
\begin{gather*}
( gb,\psi ) \sim ( g,\lambda ( b) \psi ) .
\end{gather*}
(The equivalence class of $( g,\varphi ) $ is denoted by: $\left[
g,\varphi \right] $.) Here, $g\in G^{c}$, $b\in B$, $\varphi \in \mathbb{C}$, $\lambda \in \mathcal{C}$, and~$\lambda ( b) $ is the character
representation $B\longrightarrow \mathbb{C}^{\times }$, def\/ined by:
\begin{gather*}
\lambda ( \exp ( iH) ) =\exp ( i\left\langle
\lambda ,H\right\rangle ) , \qquad  H\in \mathfrak{t}, \\
\lambda ( n) =1, \qquad n\in N.
\end{gather*}
These representations (for all $\lambda \in \mathcal{W}$) exhaust all one dimensional representations of $B$, since $B$ and $T$ have the same fundamental group and any one dimensional representation of $B$ must be trivial on $N$.

According to the Borel--Weil theorem \cite{Hurt-1}: if $\lambda $ is dominant,
then the $G$ action on $\mathcal{L}^{\lambda }$, def\/ined~by:
\begin{gather*}
g_{1}\cdot \left[ g,\varphi \right] =\left[ g_{1}g,\varphi \right],
\end{gather*}
is irreducible on the space of holomorphic sections $\Gamma _{\rm hol}(
\mathcal{L}^{\lambda }) $, and the corresponding representation is
equivalent to the representation $\pi _{\lambda }$ of highest weight $\lambda $ of $G$. Berezin quantization provides a method for realizing
$\pi _{\lambda }$ on $\Gamma _{\rm hol}( \mathcal{L}^{\lambda }) $
\cite{Marinov-1}. The representation space in Berezin theory is realized as a reproducing
kernel Hilbert space. Specif\/ically, let $\psi :G/T\longrightarrow $ $%
\mathcal{L}^{\lambda }$ be a holomorphic section of $\mathcal{L}^{\lambda }$, let $( z,\psi ( z) ) $ be the trivialization of $\psi $ on some open neighborhood $U$. Then the inner product on the
space of sections in the Berezin theory is def\/ined by:
\begin{gather*}
( \psi _{1},\psi _{2}) =\frac{N^{\lambda }}{V}\int_{G/T}\psi
_{1}( z) \overline{\psi _{2}( z) }\exp \big({-}K^{( \lambda ) }( z,\overline{z}) \big) d\mu (
z,\overline{z}),
\end{gather*}
where the sum over an appropriate open cover of $G/T$ weighted by a partition of unity \cite{Peetre-1} is implicit. Here, $\psi _{1},\psi _{2}\in L^{2}( \Gamma
_{\rm hol}( \mathcal{L}^{\lambda }) ) $, $N^{\lambda }$ is the
dimension of the representation $\pi _{\lambda }$, and $V$ is the total mass
of the Liouville measure $d\mu ( z,\overline{z}) $. We note that due to the compatibility of the connection on $\mathcal{L}^{\lambda }$ def\/ined by the K\"ahler potential and the Hermitian metric on the Fibers, all the integrations over $G/T$ in this work are of global functions.

The representatives of group elements and of elements of the universal enveloping
algebra of its Lie algebra are realized in the Berezin theory by means of
symbols, which act as integration kernels on $L^{2}( \Gamma
_{\rm hol}( \mathcal{L}^{\lambda }) ) $. The action of an
operator $A$ in the representation space of $\pi _{\lambda}$on $%
L^{2}( \Gamma _{\rm hol}( \mathcal{L}^{\lambda }) ) $ is
realized by means of its symbol $A^{\lambda }( \zeta ,\overline{z}%
) :$ $G/T\times G/T\longrightarrow \mathbb{C}$ as follows:
\begin{gather*}
( \pi _{\lambda}( A) \circ \psi ) (
\zeta )
=\frac{N^{\lambda }}{V}\int_{G/T}A^{\lambda }( \zeta ,\overline{z}
) \exp \big( K^{( \lambda ) }( \zeta ,\overline{z}
) \big) \psi ( z) \exp \big({-}K^{( \lambda )
}( z,\overline{z}) \big) d\mu ( z,\overline{z}).
\end{gather*}

We def\/ine the covariant principal Berezin symbol of $A$ by:
\begin{gather*}
\widetilde{A}^{\lambda }( \zeta ,\overline{z}) =A^{\lambda
}( \zeta ,\overline{z}) \exp \big( K^{( \lambda )
}( \zeta ,\overline{z}) \big).
\end{gather*}
In the parametrization we use, the principal covariant symbols have the property
that their~re\-striction to the largest Schubert cell $\Sigma _{s}\subset
G/T$ is polynomial in the af\/f\/ine coordinates of~$\Sigma _{s}$~\cite{Marinov-1, Marinov-2}.

The reproduction property of is expressed through the relation:
\begin{gather*}
\psi ( \zeta ) =\frac{N^{\lambda }}{V}\int_{G/T}L^{\lambda
}( \zeta ,\overline{z}) \psi ( z) \exp \big(
{-}K^{( \lambda ) }( z,\overline{z}) \big) d\mu (
z,\overline{z}) ,
\end{gather*}
where $L^{\lambda }( \zeta ,\overline{z}) $ is the reproducing
kernel of $L^{2}( \Gamma _{\rm hol}( \mathcal{L}^{\lambda })
) $. The reproducing kernel can be viewed as the principal covariant
symbol of the unit operator in $L^{2}( \Gamma _{\rm hol}( \mathcal{L}^{\lambda })) $. On the other hand
it is the kernel of the orthogonal projector on $L^{2}( \Gamma _{\rm hol}( \mathcal{L%
}^{\lambda }) ) $ in $L^{2}( \Gamma ( \mathcal{L}%
^{\lambda }) ) $.

Clearly:
\begin{gather*}
L^{\lambda }( \zeta ,\overline{z}) =\sum\limits_{j=1}^{N^{\lambda
}}\widehat{\psi }_{j}( \zeta ) \overline{\widehat{\psi }%
_{j}( z) },
\end{gather*}
where $\big\{ \widehat{\psi _{j}}, \; j=1,\ldots ,N^{\lambda }\big\}
$ is any orthonormal set of $L^{2}( \Gamma _{\rm hol}( \mathcal{L}
^{\lambda }) ) $. We choose to work in the weight basis in which the basic vectors of the orthonormal set are indexed by their weight vectors~$m$ corresponding to the Cartan generators and the corresponding degeneracy index~$i_{m}$, in terms of which the latter equation can be rephrased as:
\begin{gather}
L^{\lambda }( \zeta ,\overline{z}) =\sum\limits_{m\in W^{\lambda
}}\sum\limits_{i_{m}=1}^{\gamma ^{\lambda }( m) }\widehat{\psi}
_{m,i_{m}}( \zeta ) \overline{\widehat{\psi }_{m,
i_{m}}( z)},
\label{eqn:RK}
\end{gather}
where $W^{\lambda }$ is the weight set of $\pi _{\lambda }$ and $\gamma ^{\lambda }( m) $ is the multiplicity of the weight $m$
in $\pi _{\lambda}$. A~basic property of $L^{2}( \Gamma
_{\rm hol}( \mathcal{L}^{\lambda }) ) $, as a reproducing
kernel Hilbert space, is that its reproducing kernel is given in terms of
the analytic continuation of the corresponding K\"ahler potential by the
relation:
\begin{gather}
L^{\lambda }( \zeta ,\overline{z}) =\exp \big( K^{\lambda
}( \zeta ,\overline{z}) \big).
\label{eqn:LeqExpK}
\end{gather}

In Berezin's original work \cite{Berezin-1} the condition (\ref{eqn:LeqExpK}) was imposed in order to establish a quantization (correspondence principle). In the case under study of the reproducing kernel Hilbert spaces of holomorphic sections of line bundles over compact K\"ahler manifolds, the validity of~(\ref{eqn:LeqExpK}) was established in~\cite{Rawnsley-1}.

We shall refer to the K\"ahler potentials $\left\{ K^{j}( \zeta ,%
\overline{z}) ,\,j=1,\dots ,\mathrm{rank}( \mathfrak{g})
\right\} $ corresponding the fundamental tensor representations, i.e., the
representations of highest weights equal to fundamental weights $\left\{
w_{j}\in \mathcal{W},\ j=1,\dots ,\mathrm{rank}( \mathfrak{g})
\right\} $, as the basic K\"ahler potentials. The corresponding reproducing
kernels will be addressed to as the basic reproducing kernels:
\begin{gather*}
L^{j}( \zeta ,\overline{z}) \equiv L^{w_{j}}( \zeta ,%
\overline{z}) =\exp ( K^{j}( \zeta ,\overline{z})
).
\end{gather*}

Therefore, the K\"ahler potential corresponding to a general highest weight
representation is given by:
\begin{gather*}
 K^{\lambda }( \zeta ,\overline{z}) =\sum\limits_{j=1}^{\mathrm{rank}
( \mathfrak{g}) }l_{j}K^{j}( \zeta ,\overline{z}) , \qquad
\lambda =\sum\limits_{j=1}^{\mathrm{rank}( \mathfrak{g}) }l_{j}w_{j}, \qquad
l_{j}\in \mathbb{N\cup }\left\{ 0\right\} ,\quad  j=1,\ldots ,\mathrm{rank}(
\mathfrak{g}).
\end{gather*}

\subsection[The $T$ action on $G/T$ and $L^{2}( \Gamma _{\rm hol}(
\mathcal{L}^{\lambda }) )$]{The $\boldsymbol{T}$ action on $\boldsymbol{G/T}$ and $\boldsymbol{L^{2}( \Gamma _{\rm hol}(
\mathcal{L}^{\lambda }) )}$}

$G/T$ can be parametrized through the canonical dif\/feomorphism $G/T\cong
G^{c}/B $ by a holomorphic section $\xi ( z) :$ $%
G^{c}/B\longrightarrow G^{c}$ of the principal bundle $B\longrightarrow
G^{c}\longrightarrow $ $G^{c}/B$ . The $T$ action on $G/T$ is given by:
\begin{gather*}
\xi ( h\cdot z) =h\xi ( z) h^{-1}, \qquad  h\in T.
\end{gather*}

The induced $T$ action on $L^{2}( \Gamma _{\rm hol}( \mathcal{L}%
^{\lambda }) ) $ can be obtained from the $\mathcal{L}%
^{\lambda }$ property as a homogeneous bundle:
\begin{gather}
( h\circ \psi ) ( z) =\lambda ( h) \psi
( h^{-1}\cdot z).
\label{eqn:Taction}
\end{gather}

\subsection[Parametrization of $G/T$ and the construction of the basic
K\"ahler potentials]{Parametrization of $\boldsymbol{G/T}$ and the construction\\ of the basic
K\"ahler potentials}

In this section and in the computational examples, all group and universal
enveloping algebra elements of $G^{c}$ will be represented in the
basic fundamental representation of $G$ (i.e., operators in the
representation space of the basic fundamental representation of $G$).

We use the Bando--Kuratomo--Maskawa--Uehera method \cite{Bando-1, Bando-2} for the construction of the basic K\"ahler potentials.
Let $\left\{ Y_{j}\in \mathfrak{t}, \; j=1,\ldots ,\mathrm{rank}( \mathfrak{g}%
) \right\} $ be the set of coweights of $\mathfrak{t}$ def\/ined by:
\begin{gather*}
\left\langle \alpha ,Y_{j}\right\rangle =\sum\limits_{k=1}^{\mathrm{rank}%
( \mathfrak{g}) }G_{jk}\, w_{k}\cdot \alpha ,
\end{gather*}

\noindent where $G_{jk}=( A^{-1}) _{ij}\frac{\gamma ^{( i)
}\cdot \gamma ^{( j) }}{2}$ is metric on the weight space in the
weight basis ($A$ is the Cartan matrix). In the terminology of \cite{Bando-1} and \cite{Bando-2}, the generators $Y_{j}$ are called central charges. Let $\left\{ \eta _{j}\in \mathfrak{t}, \; j=1,\ldots ,
\mathrm{rank}( \mathfrak{g}) \right\} $ be the set of projectors on the
lowest value eigenspace of $Y_{j}$ (in the basic fundamental representation). Then
according to \cite{Bando-1} and \cite{Bando-2}, the basic K\"ahler potentials can be constructed
according to:
\begin{gather*}
K^{j}( \zeta ,\overline{z}) =\log \big( \det \big( \eta _{j}\xi
( z) ^{\dag}\xi ( \zeta ) \eta _{j}+1-\eta _{j}\big)
\big) , \qquad  j=1,\ldots ,\mathrm{rank}( \mathfrak{g}) ,
\end{gather*}
where $( \cdot ) ^{\dag}$ denotes Hermitian conjugation.

For the specif\/ic computations, we use the parametrization of $G/T$, obtained
from the holomorphic dif\/feomorphism: $N_{-}\longrightarrow \Sigma _{s},$
given by:
\begin{gather}
\xi ( z) =\exp \Bigg( \sum\limits_{\alpha \in \Delta
^{+}}z^{\alpha }E_{\alpha }\Bigg) , \qquad z^{\alpha }\in \mathbb{C}.
\label{eqn:parameterize}
\end{gather}

\section{A weight multiplicity formula based on Berezin quantization}\label{WMF}

The action on $L^{2}( \Gamma _{\rm hol}( \mathcal{L}^{\lambda })
) $ can be used to project the representation space onto the subspace
spanned by a given weight $m$, as follows.

On one hand, since $\widehat{\psi }_{m,j_{m}}$ is a section
corresponding to a vector of weight $m$, then:
\begin{gather*}
\big( h\circ \widehat{\psi }_{m,j_{m}}\big) ( z)
=m( h) \widehat{\psi }_{m,j_{m}}( z).
\end{gather*}

On the other hand, according to (\ref{eqn:Taction}):
\begin{gather*}
\big( h\circ \widehat{\psi}_{m,j_{m}}\big) ( z)
=\lambda ( h) \widehat{\psi }_{m,j_{m}}( h\cdot z).
\end{gather*}

Combining the two equations, we obtain:
\begin{gather*}
\widehat{\psi }_{m,j_{m}}( h\cdot z) =( \lambda -m)
( h) \widehat{\psi }_{m,j_{m}}( z).
\end{gather*}

Thus, the projection onto the subspace of a given weight $m$, can be
obtained by the following integration over $T$:
\begin{gather*}
\int_{T}( m-\lambda ) ( h) \widehat{\psi }_{m',j_{m'}}( h\cdot z) d\mu ( h) =\delta _{m,m^{\prime }}
\widehat{\psi }_{m,j_{m}}( z),
\end{gather*}
where $d\mu ( h) $ is the Lebesgue measure on $T$. Applying this
projection operation to the reproducing kernel $L^{\lambda }( \zeta ,\overline{z}) $, we obtain:
\begin{gather*}
\sum\limits_{i_{m}=1}^{\gamma ^{\lambda }( m) }\widehat{\psi }_{m,%
 i_{m}}( \zeta ) \overline{\widehat{\psi }_{m,
i_{m}}( z) } =
  \int_{T}( m-\lambda ) ( h) \Bigg( \sum\limits_{m\in
W^{\lambda }}\sum\limits_{i_{m}=1}^{\gamma ^{\lambda }( m) }
\widehat{\psi }_{m,i_{m}}( h\cdot \zeta ) \overline{\widehat{%
\psi }_{m,i_{m}}( z) }\Bigg) d\mu ( h) ,
\end{gather*}
which, upon using (\ref{eqn:RK}) and (\ref{eqn:LeqExpK}), is the statement of Proposition~\ref{prop-2}:
\begin{gather}
L_{m}^{\lambda }( \zeta ,\overline{z}) =\int_{T}( m-\lambda
) ( h) \exp \big( K^{\lambda }( h\cdot \zeta ,\overline{z}
) \big) d\mu ( h) ,
\label{eqn:projEqn}
\end{gather}
 where
\begin{gather}
L_{m}^{\lambda }( \zeta ,\overline{z})
=\sum\limits_{i_{m}=1}^{\gamma ^{\lambda }( m) }\widehat{\psi }
_{m, i_{m}}( \zeta ) \overline{\widehat{\psi }_{m,
i_{m}}( z) }.
\label{eqn:projPsi}
\end{gather}

The principal symbol $L_{m}^{\lambda }( \zeta ,\overline{z}) $\
of the projector in  $L^{2}( \Gamma _{\rm hol}( \mathcal{L}^{\lambda }) )$ onto the subspace spanned by vectors of weight $m$ acts
as a reproducing kernel on the same subspace. Now, since the sections~$\widehat{\psi }_{m, i_{m}}$ are orthonormal, we get:
\begin{gather}
\gamma ^{\lambda }( m) =\frac{N^{\lambda }}{V}\int_{G/T}L_{m}^{%
\lambda }( \zeta ,\overline{z}) \exp \big( {-}K^{( \lambda
) }( z,\overline{z}) \big) d\mu ( z,\overline{z}
) .
\label{eqn:multipEqn}
\end{gather}

Combining (\ref{eqn:projEqn}) with (\ref{eqn:multipEqn}) we obtain the statement of Proposition~\ref{prop-1}:
\begin{gather*}
\gamma ^{\lambda }( m) =\frac{N^{\lambda }}{V}\int_{G/T}\int_{T}%
( m-\lambda ) ( h) \exp \big( K^{\lambda }(
h\cdot \zeta ,\overline{z}) \big) \exp \big( {-}K^{\lambda}( z,\overline{z}) \big) d\mu ( h) d\mu ( z,
\overline{z}).
\end{gather*}

\section{A method for weight multiplicity computation}\label{WMcomp}

\subsection{Computation by direct integration}

The weight multiplicity formula given in the previous section is not
constructive. While the construction of the K\"ahler potentials along the
method described in Section~\ref{Ber-BW} and the integration over the maximal torus are straightforward, no simple rules are known in advance for the integration of
holomorphic sections. Although, the integration can be performed by
elementary integration techniques on the largest Schubert cell, since its
complement in $G/T$ is of Liouville measure zero, but even for spaces as
small as $Fl( 3) \cong SU( 3)/S( U( 1)\times U( 1) \times U( 1) ) $ and small representations, direct
integration is a quite lengthy task.

\subsection{Weight multiplicity computation as a rank of a Hermitian form}

The decomposition of $L_{m}^{\lambda }( \zeta ,\overline{z}) $
given in (\ref{eqn:projPsi}) allows multiplicity computations without direct integration. In the parametrization of $G/T$ def\/ined in (\ref{eqn:parameterize}), the restriction of any principal Berezin symbol to the largest Schubert cell $\Sigma _{s}$ is polynomial in the af\/f\/ine coordinates $( \zeta ^{\alpha },\overline{z}%
^{\alpha }) $. Let $d^{\alpha }$ be the polynomial degree of
coordinate $z^{\alpha }$ in the restriction of $L_{m}^{\lambda }( \zeta
,\overline{z}) $, and def\/ine the monomials:
\begin{gather}
u_{n}( \zeta ) =\prod\limits_{\alpha \in \Delta ^{+}}( \zeta
^{\alpha }) ^{n_{\alpha }},  \qquad  v_{n}( z)
=\prod\limits_{\alpha \in \Delta ^{+}}( z^{\alpha }) ^{n_{\alpha
}}, \qquad  0\leq n_{\alpha }\leq d_{\alpha }, \nonumber \\
n=( n_{\alpha _{1}},n_{\alpha _{2}},\dots, n_{\alpha _{D}}) ,  \qquad
D=\dim_\mathbb{C} G/T .
 \label{eqn:pol}
\end{gather}

Clearly, any section of the orthonormal subset of weight $m$ is a linear
combination of the monomials of (\ref{eqn:pol}):
\begin{gather}
\widehat{\psi }_{m,i_{m}}( \zeta )
=\sum\nolimits_{n}a_{i_{m}}^{n}u_{n}( \zeta ).
\label{eqn:psiPol}
\end{gather}

 The monomials in (\ref{eqn:pol}), def\/ine a biholomorphic transformation $
f:\Sigma _{s}\times \Sigma _{s}\longrightarrow \mathbb{V}\times  \mathbb{V}
$, $\mathbb{V\cong C}^{d} $, $d= \sum_{\alpha \in \Delta
^{+}}d_{\alpha }$. Consider the Hermitian form on $\mathbb{V}$ def\/ined by:
\begin{gather*}
\mathbf{L}_{m}^{\lambda }( u,\overline{v})
=\sum\limits_{i_{m}=1}^{\gamma ^{\lambda }( m) }\Big(
\sum\nolimits_{n}a_{i_{m}}^{n}u_{n}\Big) \Big( \sum\nolimits_{n}\overline{
a}_{i_{m}}^{n}\overline{v}_{n}\Big).
\end{gather*}
 We have: $f\ast \mathbf{L}_{m}^{\lambda }=L_{m}^{\lambda }$.
Suppose there exists another Hermitian form $\mathbf{L}_{m}^{\prime \lambda
}$ such that $f\ast \mathbf{L}_{m}^{\prime \lambda }=L_{m}^{\lambda }$.
This would imply that: $f\ast ( \mathbf{L}_{m}^{\lambda }-\mathbf{L}%
_{m}^{\prime \lambda }) =0$. But $f\ast ( \mathbf{L}_{m}^{\lambda
}-\mathbf{L}_{m}^{\prime \lambda }) $ is a polynomial function on $\Sigma _{s}\times \Sigma _{s}$, therefore it can be identically $0$ only if
all of its coef\/f\/icients vanish. This would imply that the coef\/f\/icients of
the Hermitian form $\mathbf{L}_{m}^{\lambda }-\mathbf{L}_{m}^{\prime \lambda
}$ vanish, hence $\mathbf{L}_{m}^{\prime \lambda }=\mathbf{L}_{m}^{\lambda }$.

Let $A$ be the matrix of dimension $\gamma ^{\lambda }( m) \times
d$ whose coef\/f\/icients are $a_{i_{m}}^{n}$. Clearly $d\geq \gamma ^{\lambda
}( m) $, and also the row vectors of $A$ are linearly
independent, otherwise, at least one of the sections in (\ref{eqn:psiPol}) is a linear
combination of the others, which is impossible, since they are orthonormal.
Therefore,
\begin{gather*}
\mathrm{rank}( A) =\gamma ^{\lambda }( m).
\end{gather*}

The Hermitian matrix of the Hermitian form $\mathbf{L}_{m}^{\lambda }$ is $A^{\dag}A$, whose rank is identical to $A$, hence we obtained the proof of
Proposition~\ref{prop-3}:
\begin{gather*}
\gamma ^{\lambda }( m) =\mathrm{rank}_{\mathbb{V}}\big( \mathbf{L}_{m}^{\lambda }\big).
\end{gather*}

\appendix

\section{Examples}\label{Exmps}

\subsection[Computation of some weight multiplicities in the representation $
\pi _{(4,2)}$ of $SU(3)$]{Computation of some weight multiplicities\\ in the representation $
\boldsymbol{\pi _{(4,2)}}$ of $\boldsymbol{SU(3)}$}

A suitable basis for the generators of $\mathfrak{sl}(3)=\mathfrak{su}(3)^{c}$ in
the basic fundamental representation is given by:

Positive root generators:
\begin{gather*}
E_{\alpha _{1}}=
\left(
\begin{array}{ccc}
0 & 1 & 0 \\
0 & 0 & 0 \\
0 & 0 & 0
\end{array}
\right) , \qquad E_{\alpha _{2}}=\left(
\begin{array}{ccc}
0 & 0 & 0 \\
0 & 0 & 1 \\
0 & 0 & 0
\end{array}
\right) , \qquad E_{\alpha _{3}}=\left(
\begin{array}{ccc}
0 & 0 & 1 \\
0 & 0 & 0 \\
0 & 0 & 0
\end{array}
\right).
\end{gather*}

Negative root generators:
\begin{gather*}
E_{-\alpha _{1}}=E_{\alpha _{1}}^{\dag}, \qquad E_{-\alpha _{2}}=E_{-\alpha
_{2}}^{\dag}, \qquad E_{-\alpha _{3}}=E_{\alpha _{3}}^{\dag}.
\end{gather*}

Cartan subalgebra generators:
\begin{gather*}
H_{1}=\left[ E_{\alpha _{1}},E_{-\alpha _{1}}\right] =\mathrm{diag}(
1,-1,0) \qquad H_{2}=\left[ E_{\alpha _{2}},E_{-\alpha _{2}}\right] =\mathrm{diag}( 0,1,-1).
\end{gather*}

Central charges:
\begin{gather*}
Y_{1}=G_{11}H_{1}+G_{12}H_{2}=\tfrac{1}{3}\mathrm{diag}( 2,-1,-1),\qquad
Y_{2}=G_{21}H_{1}+G_{22}H_{2}=\tfrac{1}{3}\mathrm{diag}( 1,1,-2).
\end{gather*}

The projectors required:
\begin{gather*}
\eta _{1}=\mathrm{diag}( 0,1,1), \qquad \eta _{2}=\mathrm{diag}(
0,0,1).
\end{gather*}

The coset representative:
\begin{gather*}
\xi ( z) =\exp \left( \sum\limits_{i=1}^{3}z_{i}E_{\alpha
_{i}}\right) =\left(
\begin{array}{ccc}
1 & z_{1} & z_{3}-z_{1}z_{2}/2 \\
0 & 1 & z_{2} \\
0 & 0 & 1
\end{array}
\right).
\end{gather*}

The action of the element $h=\exp ( i\theta _{1}H_{1}+i\theta
_{2}H_{2}) \in T$ on the af\/f\/ine coordinates:
\begin{gather*}
( h\cdot z) _{1}=\exp ( 2i\theta _{1}-i\theta _{2}) z_{1},
\qquad ( h\cdot z) _{2}=\exp ( -i\theta _{1}+2i\theta _{2})
z_{2}, \\  ( h\cdot z) _{3}=\exp ( i\theta _{1}+i\theta _{2})
z_{3}.
\end{gather*}

The fundamental reproducing kernels:
\begin{gather*}
L^{1}( \zeta ,\overline{z}) =\det \big( \eta _{1}\xi (
z) ^{\dag}\xi ( \zeta ) \eta _{1}+1-\eta _{1}\big) =1+\zeta
_{1}\overline{z}_{1}+\zeta _{3}^{+}\overline{z}_{3}^{+}, \\
L^{2}( \zeta ,\overline{z}) =\det \big( \eta _{2}\xi (
z) ^{\dag}\xi ( \zeta ) \eta _{2}+1-\eta _{2}\big) =1+\zeta
_{2}\overline{z}_{2}+\zeta _{3}^{-}\overline{z}_{3}^{-},
\end{gather*}
where $\zeta _{3}^{\pm }=\zeta _{3}\pm \zeta _{1}\zeta _{2}/2$, and
similarly for the antiholomorphic coordinates.

The reproducing kernel of $L^{2}( \Gamma _{\rm hol}( \mathcal{L}%
^{( 4,2) }) ) $:
\begin{gather*}
L^{( 4,2) }( \zeta ,\overline{z}) =( 1+\zeta _{1}%
\overline{z}_{1}+\zeta _{3}^{+}\overline{z}_{3}^{+}) ^{4}(
1+\zeta _{2}\overline{z}_{2}+\zeta _{3}^{-}\overline{z}_{3}^{-}) ^{2}.
\end{gather*}

\subsubsection[Computation of the multiplicity of the weight $(
0,1) $ in $ \pi _{(4,2)}$]{Computation of the multiplicity of the weight $\boldsymbol{(
0,1)}$ in $\boldsymbol{\pi _{(4,2)}}$}

Computation of the projector onto the weight space $( -2,-1) $ in
$L^{2}( \Gamma _{\rm hol}( \mathcal{L}^{( 4,2) })
) $:
\begin{gather*}
L_{( -2,-1) }^{( 4,2) }( \zeta ,\overline{z}
)   =\frac{1}{( 2\pi ) ^{2}}\int\limits_{0}^{2\pi }d\theta
_{1}\int\limits_{0}^{2\pi }d\theta _{2}\exp ( -6i\theta _{1}-4i\theta
_{2}) \\
\phantom{L_{( -2,-1) }^{( 4,2) }( \zeta ,\overline{z})   =}{}
 \times L^{( 4,2) }( \exp ( 2i\theta _{1}-i\theta _{2})
\zeta _{1},\exp ( -i\theta _{1}+2i\theta _{2}) \zeta _{2},\exp
( i\theta _{1}+i\theta _{2}) \zeta _{3},\overline{z}_{1},%
\overline{z}_{2},\overline{z}_{3}).
\end{gather*}

The integration result:
\begin{gather*}
L_{( -2,-1) }^{( 4,2) }( \zeta ,\overline{z})   =
\tfrac{17}{64}\zeta _{1}^{5}\zeta _{2}^{4}\overline{z}_{1}^{5}%
\overline{z}_{2}^{4}+6\zeta _{1}^{3}\zeta _{2}^{2}\zeta _{3}^{2}\overline{z}%
_{1}^{3}\overline{z}_{2}^{2}\overline{z}_{3}^{2}+20\zeta _{1}\zeta _{3}^{4}%
\overline{z}_{1}\overline{z}_{3}^{4}+\tfrac{5}{4}\zeta _{1}^{4}\zeta
_{2}^{3}\zeta _{3}\overline{z}_{1}^{4}\overline{z}_{2}^{3}\overline{z}_{3}%
 \\
\phantom{L_{( -2,-1) }^{( 4,2) }( \zeta ,\overline{z})=}{} +20\zeta _{1}^{2}\zeta _{2}\zeta _{3}^{3}\overline{z}_{1}^{2}\overline{z}_{2}%
\overline{z}_{3}^{3}  +\tfrac{3}{2}\zeta _{1}^{4}\zeta _{2}^{3}\zeta _{3}\overline{z}_{1}^{3}%
\overline{z}_{2}^{2}\overline{z}_{3}^{2}+\tfrac{3}{2}\zeta _{1}^{3}\zeta
_{2}^{2}\zeta _{3}^{2}\overline{z}_{1}^{4}\overline{z}_{2}^{3}\overline{z}_{3}%
+2\zeta _{1}^{2}\zeta _{2}\zeta _{3}^{3}\overline{z}_{1}^{5}\overline{z}%
_{2}^{4} \\
\phantom{L_{( -2,-1) }^{( 4,2) }( \zeta ,\overline{z})=}{}
+2\zeta _{1}^{5}\zeta _{2}^{4}\overline{z}_{1}^{2}\overline{z}_{2}%
\overline{z}_{3}^{3}  -\tfrac{1}{2}\zeta _{1}^{4}\zeta _{2}^{3}\zeta _{3}\overline{z}_{1}^{5}%
\overline{z}_{2}^{4}-\tfrac{1}{2}\zeta _{1}^{5}\zeta _{2}^{4}\overline{z}%
_{1}^{4}\overline{z}_{2}^{3}\overline{z}_{3}-5\zeta _{1}^{4}\zeta _{2}^{3}\zeta
_{3}\overline{z}_{1}^{2}\overline{z}_{2}\overline{z}_{3}^{3} \\
\phantom{L_{( -2,-1) }^{( 4,2) }( \zeta ,\overline{z})=}{}
-5\zeta
_{1}^{2}\zeta _{2}\zeta _{3}^{3}\overline{z}_{1}^{4}\overline{z}_{2}^{3}%
\overline{z}_{3}  -6\zeta _{1}^{2}\zeta _{2}\zeta _{3}^{3}\overline{z}_{1}^{3}\overline{z}%
_{2}^{2}\overline{z}_{3}^{2}-6\zeta _{1}^{3}\zeta _{2}^{2}\zeta _{3}^{2}%
\overline{z}_{1}^{2}\overline{z}_{2}\overline{z}_{3}^{3}-6\zeta _{1}\zeta
_{3}^{4}\overline{z}_{1}^{3}\overline{z}_{2}^{2}\overline{z}_{3}^{2} \\
\phantom{L_{( -2,-1) }^{( 4,2) }( \zeta ,\overline{z})=}{}
-6\zeta
_{1}^{3}\zeta _{2}^{2}\zeta _{3}^{2}\overline{z}_{1}\overline{z}_{3}^{4}
  +2\zeta _{1}\zeta _{3}^{4}\overline{z}_{1}^{4}\overline{z}_{2}^{3}%
\overline{z}_{3}+2\zeta _{1}^{4}\zeta _{2}^{3}\zeta _{3}\overline{z}_{1}%
\overline{z}_{3}^{4}+\tfrac{1}{4}\zeta _{1}\zeta _{3}^{4}\overline{z}_{1}^{5}%
\overline{z}_{2}^{4}+\tfrac{1}{4}\zeta _{1}^{5}\zeta _{2}^{4}\overline{z}_{1}%
\overline{z}_{3}^{4} \\
 \phantom{L_{( -2,-1) }^{( 4,2) }( \zeta ,\overline{z})=}{} -8\zeta _{1}^{2}\zeta _{2}\zeta _{3}^{3}\overline{z}_{1}\overline{z}%
_{3}^{4}-8\zeta _{1}\zeta _{3}^{4}\overline{z}_{1}^{2}\overline{z}_{2}%
\overline{z}_{3}^{3}-\tfrac{9}{8}\zeta _{1}^{3}\zeta _{2}^{2}\zeta _{3}^{2}%
\overline{z}_{1}^{5}\overline{z}_{2}^{4}-\tfrac{9}{8}\zeta _{1}^{5}\zeta
_{2}^{4}\overline{z}_{1}^{3}\overline{z}_{2}^{2}\overline{z}_{3}^{2}
\end{gather*}

The monomial set of $L_{( -2,-1) }^{( 4,2) }$:
\begin{gather*}
u_{1}=\zeta _{1}\zeta _{3}^{4}, \qquad u_{2}=\zeta _{1}^{2}\zeta _{2}\zeta
_{3}^{3}, \qquad u_{3}=\zeta _{1}^{3}\zeta _{2}^{2}\zeta _{3}^{2}, \qquad u_{4}=\zeta
_{1}^{4}\zeta _{2}^{3}\zeta _{3}, \qquad u_{5}=\zeta _{1}^{5}\zeta _{2}^{4}, \\
v_{1}=z_{1}z_{3}^{4}, \qquad v_{2}=z_{1}^{2}z_{2}z_{3}^{3}, \qquad
v_{3}=z_{1}^{3}z_{2}^{2}z_{3}^{2}, \qquad v_{4}=z_{1}^{4}z_{2}^{3}z_{3}, \qquad
v_{5}=z_{1}^{5}z_{2}^{4}.
\end{gather*}

The Hermitian form $\mathbf{L}_{( -2,-1) }^{(
4,2) }( u,\overline{v}) $:
\begin{gather*}
\mathbf{L}_{( -2,-1) }^{( 4,2) }( u,\overline{v}%
) =\left(
\begin{array}{c}
v_{1} \\
v_{2} \\
v_{3} \\
v_{4} \\
v_{5}
\end{array}
\right) ^{\dag}\left(
\begin{array}{ccccc}
20 & -8 & -6 & 2 & \frac{1}{4} \vspace{1mm}\\
-8 & 20 & -6 & -5 & 2 \vspace{1mm}\\
-6 & -6 & 6 & \frac{3}{2} & \frac{-9}{8} \vspace{1mm}\\
2 & -5 & \frac{3}{2} & \frac{5}{4} & \frac{-1}{2} \vspace{1mm}\\
\frac{1}{4} & 2 & \frac{-9}{8} & \frac{-1}{2} & \frac{17}{64}
\end{array}
\right) \left(
\begin{array}{c}
u_{1} \\
u_{2} \\
u_{3} \\
u_{4} \\
u_{5}
\end{array}
\right).
\end{gather*}

Computation of the multiplicity:
\begin{gather*}
\gamma ^{( 4,2) }( ( -2,-1) ) =\mathrm{rank}%
\big( \mathbf{L}_{( -2,-1) }^{( 4,2) }( u,\overline{v}) \big) =2.
\end{gather*}

\subsubsection[Computation of the multiplicity of the weight $(-6,4) $ in $\pi _{(4,2)}$]{Computation of the multiplicity of the weight $\boldsymbol{(-6,4)}$ in $\boldsymbol{\pi _{(4,2)}}$}

This weight lies on the Weyl group orbit of the highest weight, therefore
its multiplicity should be 1. Repeating the same type of computation as in
the previous case, we obtain:
\begin{gather*}
L_{( -6,4) }^{( 4,2) }( \zeta ,\overline{z})  =
\tfrac{1}{256}\zeta _{1}^{4}\zeta _{3}^{2}\overline{z}_{1}^{4}%
\overline{z}_{3}^{2}+\tfrac{1}{64}\zeta _{1}^{5}\zeta _{2}\zeta _{3}\overline{z}_{1}^{5}\overline{z}_{2}\overline{z}_{3}+\tfrac{1}{256}\zeta _{1}^{6}\zeta
_{2}^{2}\overline{z}_{1}^{6}\overline{z}_{2}^{2}+\tfrac{1}{128}\zeta _{1}^{5}\zeta _{2}\zeta _{3}\overline{z}_{1}^{4}%
\overline{z}_{3}^{2}\\
\phantom{L_{( -6,4) }^{( 4,2) }( \zeta ,\overline{z})  =}{}
 + \tfrac{1}{256}\zeta _{1}^{6}\zeta _{2}^{2}%
\overline{z}_{1}^{4}\overline{z}_{3}^{2}+\tfrac{1}{128}\zeta _{1}^{6}\zeta
_{2}^{2}\overline{z}_{1}^{5}\overline{z}_{2}\overline{z}_{3}
 +\tfrac{1}{128}\zeta _{1}^{4}\zeta _{3}^{2}\overline{z}_{1}^{5}\overline{z}_{2}%
\overline{z}_{3}+\tfrac{1}{256}\zeta _{1}^{4}\zeta _{3}^{2}%
\overline{z}_{1}^{6}\overline{z}_{2}^{2}\\
\phantom{L_{( -6,4) }^{( 4,2) }( \zeta ,\overline{z})  =}{}
+\tfrac{1}{128}\zeta _{1}^{5}\zeta
_{2}\zeta _{3}\overline{z}_{1}^{6}%
\overline{z}_{2}^{2}.
\end{gather*}

The monomial set of $L_{( -6,4) }^{( 4,2) }$:
\begin{gather*}
u_{1}=\zeta _{1}^{4}\zeta _{3}^{2}, \qquad u_{2}=\zeta _{1}^{5}\zeta _{2}\zeta
_{3}, \qquad u_{3}=\zeta _{1}^{6}\zeta _{2}^{2}, \\
v_{1}=z_{1}^{4}z_{3}^{2}, \qquad v_{2}=z_{1}^{5}z_{2}z_{3}, \qquad v_{3}=z_{1}^{6}z_{2}^{2}.
\end{gather*}

The Hermitian form $\mathbf{L}_{( -6,-4) }^{( 4,2)}( u,\overline{v}) $:
\begin{gather*}
\mathbf{L}_{( -2,-1) }^{( 4,2) }( u,\overline{v}%
) =\left(
\begin{array}{c}
v_{1} \\
v_{2} \\
v_{3}
\end{array}
\right) ^{\dag}\left(
\begin{array}{ccc}
\frac{1}{256} & \frac{1}{128} & \frac{1}{256} \vspace{1mm}\\
\frac{1}{128} & \frac{1}{64} & \frac{1}{128} \vspace{1mm}\\
\frac{1}{256} & \frac{1}{128} & \frac{1}{256}
\end{array}
\right) \left(
\begin{array}{c}
u_{1} \\
u_{2} \\
u_{3}
\end{array}
\right).
\end{gather*}

Computation of the multiplicity:
\begin{gather*}
\gamma ^{( 4,2) }( ( -6,4) ) =\mathrm{rank}%
\big( \mathbf{L}_{( -6,4) }^{( 4,2) }( u,%
\overline{v}) \big) =1.
\end{gather*}

\subsection[Computation of some weight multiplicities in the representation $\pi _{(1,1)}$ of $SO(5)$]{Computation of some weight multiplicities\\ in the representation $\boldsymbol{\pi _{(1,1)}}$ of $\boldsymbol{SO(5)}$}

We choose to work in the four dimensional basic fundamental representation of $%
\mathfrak{sp}( 2) \mathfrak{\cong so}( 5) $. We use a
quaternionic basis for the generators, with:
\begin{gather*}
\underline{1}=\left(
\begin{array}{cc}
1 & 0 \\
0 & 1
\end{array}
\right),\qquad  \sigma _{+}=\left(
\begin{array}{cc}
0 & 1 \\
0 & 0
\end{array}
\right),\qquad  \sigma _{-}=\left(
\begin{array}{cc}
0 & 0 \\
1 & 0
\end{array}
\right), \qquad \sigma _{0}=\left(
\begin{array}{cc}
1 & 0 \\
0 & -1
\end{array}
\right).
\end{gather*}

A suitable basis for the generators of $\mathfrak{sp}( 2) ^{c}$ in
the basic fundamental representation is given by:

Positive root generators:
\begin{gather*}
E_{\alpha _{1}}=\underline{1}\otimes \sigma _{+},\qquad  E_{\alpha _{2}}=\sigma
_{+}\otimes \sigma _{-}, \qquad E_{\alpha _{3}}=\sigma _{+}\otimes \sigma _{0}, \qquad
E_{\alpha _{4}}=\sigma _{+}\otimes \sigma _{+}.
\end{gather*}

Cartan subalgebra generators:
\begin{gather*}
H_{1}=\left[ E_{\alpha _{1}},E_{-\alpha _{1}}\right] =\underline{1}\otimes
\sigma _{0}, \qquad H_{2}=\left[ E_{\alpha _{2}},E_{-\alpha _{2}}\right]
=\tfrac{1}{2}( \sigma _{0}\otimes \underline{1}-\underline{1}\otimes \sigma
_{0}).
\end{gather*}

Central charges:
\begin{gather*}
Y_{1}=G_{11}H_{1}+G_{12}H_{2}=\tfrac{1}{2}( \sigma _{0}\otimes
\underline{1}+\underline{1}\otimes \sigma _{0}), \qquad
Y_{2}=G_{21}H_{1}+G_{22}H_{2}=\sigma _{0}\otimes \underline{1}.
\end{gather*}

The action of the element $h=\exp ( i\theta _{1}H_{1}+i\theta
_{2}H_{2}) \in T$ on the af\/f\/ine coordinates:
\begin{gather*}
( h\cdot z) _{1}=\exp ( 2i\theta _{1}-i\theta _{2}) z_{1},
\qquad ( h\cdot z) _{2}=\exp ( -2i\theta _{1}+2i\theta _{2})
z_{2}, \\
( h\cdot z) _{3}=\exp ( i\theta _{2}) z_{3}, \qquad (
h\cdot z) _{4}=\exp ( 2i\theta _{1}) z_{4}.
\end{gather*}

The basic reproducing kernels:
\begin{gather*}
L^{1}( \zeta ,\overline{z}) =1+\zeta _{1}\overline{z}_{1}+\zeta
_{4}^{+}\overline{z}_{4}^{+}+\zeta _{3}^{-}\overline{z}_{3}^{-}, \\
L^{2}( \zeta ,\overline{z}) =1+\zeta _{2}\overline{z}_{2}+2\zeta
_{3}^{+}\overline{z}_{3}^{+}+( \zeta _{4}^{+}-\zeta _{1}\zeta
_{3}^{+}) ( \overline{z}_{4}^{+}-\overline{z}_{1}\overline{z}%
_{3}^{+}) +( \zeta _{2}\zeta _{4}^{+}-\zeta _{3}^{-}\zeta
_{3}^{+}) ( \overline{z}_{2}\overline{z}_{4}^{+}-\overline{z}%
_{3}^{-}\overline{z}_{3}^{+}),
\end{gather*}
where $\zeta _{3}^{\pm }=\zeta _{3}\pm \zeta _{1}\zeta _{2}/2$, $\zeta
_{4}^{+}=\zeta _{4}+\zeta _{1}^{2}\zeta _{2}/6$ and similarly for the
antiholomorphic coordinates. The reproducing kernel of $L^{2}( \Gamma
_{\rm hol}( \mathcal{L}^{( 1,1) }) ) $:
\begin{gather*}
L^{( 1,1) }( \zeta ,\overline{z})  =( 1+\zeta
_{1}\overline{z}_{1}+\zeta _{4}^{+}\overline{z}_{4}^{+}+\zeta _{3}^{-}%
\overline{z}_{3}^{-}) \\
\qquad{}\times \big( 1+\zeta _{2}\overline{z}_{2}+2\zeta _{3}^{+}\overline{z}%
_{3}^{+}+( \zeta _{4}^{+}-\zeta _{1}\zeta _{3}^{+}) (
\overline{z}_{4}^{+}-\overline{z}_{1}\overline{z}_{3}^{+}) +(
\zeta _{2}\zeta _{4}^{+}-\zeta _{3}^{-}\zeta _{3}^{+}) (
\overline{z}_{2}\overline{z}_{4}^{+}-\overline{z}_{3}^{-}\overline{z}%
_{3}^{+}) \big).
\end{gather*}

The principal symbol of the projector onto the weight space $(
-1,0) $ in $L^{2}( \Gamma _{\rm hol}( \mathcal{L}^{(
1,1) }) ) $:
\begin{gather*}
L_{( -1,0) }^{( 1,1) }( \zeta ,\overline{z})   =\tfrac{7}{144}\zeta _{1}^{3}\zeta _{2}^{2}\overline{z}_{1}^{3}%
\overline{z}_{2}^{2}+\tfrac{1}{12}\zeta _{1}^{2}\zeta _{2}\zeta
_{3}  \overline{z}_{1}^{2}\overline{z}_{2}\overline{z}_{3}+3\zeta
_{3}\zeta _{4}  \overline{z}_{3}\overline{z}_{4}+2\zeta _{1}\zeta
_{3}^{2}  \overline{z}_{1}\overline{z}_{3}^{2}+\tfrac{7}{4}\zeta
_{1}\zeta _{2}  \zeta _{4}\overline{z}_{1}\overline{z}_{2}%
\overline{z}_{4} \\
\phantom{L_{( -1,0) }^{( 1,1) }( \zeta ,\overline{z})   =}{}
+\tfrac{3}{2}\zeta _{1}\zeta _{3}^{2}\overline{z}_{1}\overline{z}_{2}%
\overline{z}_{4}+\tfrac{1}{3}\zeta _{1}^{3}\zeta _{2}^{2}\overline{z}_{3}%
\overline{z}_{4}-\zeta _{1}\zeta _{3}^{2}\overline{z}_{3}\overline{z}_{4}+%
\tfrac{1}{2}\zeta _{1}\zeta _{2}\zeta _{4}\overline{z}_{3}\overline{z}_{4}-%
\tfrac{1}{12}\zeta _{1}^{3}  \zeta _{2}^{2}\overline{z}_{1}%
\overline{z}_{2}\overline{z}_{4} \\
 \phantom{L_{( -1,0) }^{( 1,1) }( \zeta ,\overline{z})   =}{}
 +\tfrac{1}{12}\zeta _{1}^{2}\zeta _{2}\zeta _{3}\overline{z}_{1}\overline{z}%
_{2}\overline{z}_{4}+\tfrac{1}{2}\zeta _{3}\zeta _{4}\overline{z}_{1}%
\overline{z}_{2}\overline{z}_{4}+\tfrac{1}{2}\zeta _{1}^{2}\zeta _{2}\zeta
_{3}\overline{z}_{3}  \overline{z}_{4}-\zeta _{3}\zeta _{4}%
\overline{z}_{1}\overline{z}_{3}^{2}-\tfrac{1}{4}\zeta _{1}^{3}\zeta _{2}^{2}%
\overline{z}_{1}\overline{z}_{3}^{2} \\
\phantom{L_{( -1,0) }^{( 1,1) }( \zeta ,\overline{z})   =}{}
  +\tfrac{1}{18}\zeta _{1}^{3}\zeta _{2}^{2}\overline{z}_{1}^{2}\overline{z}%
_{2}\overline{z}_{3}-\tfrac{1}{6}\zeta _{1}\zeta _{3}^{2}\overline{z}_{1}^{2}%
\overline{z}_{2}\overline{z}_{3}+\tfrac{1}{2}\zeta _{3}\zeta _{4}\overline{z}%
_{1}^{2}\overline{z}_{2}  \overline{z}_{3}-\tfrac{1}{6}\zeta
_{1}^{2}\zeta _{2}\zeta _{3}\overline{z}_{1}\overline{z}_{3}^{2}-\tfrac{3}{2}%
\zeta _{1}\zeta _{2}\zeta _{4}\overline{z}_{1}\overline{z}_{3}^{2} \\
\phantom{L_{( -1,0) }^{( 1,1) }( \zeta ,\overline{z})   =}{}
  -\tfrac{1}{4}  \zeta _{1}\zeta _{3}^{2}\overline{z}_{1}^{3}%
\overline{z}_{2}^{2}+\tfrac{1}{18}\zeta _{1}^{2}\zeta _{2}\zeta _{3}\overline{%
z}_{1}^{3}\overline{z}_{2}^{2}-\tfrac{1}{12}\zeta _{1}  \zeta
_{2}\zeta _{4}\overline{z}_{1}^{3}\overline{z}_{2}^{2}+\tfrac{1}{12}\zeta
_{1}\zeta _{2}\zeta _{4}\overline{z}_{1}^{2}\overline{z}_{2}\overline{z}_{3}+%
\tfrac{1}{3}\zeta _{3}\zeta _{4}  \overline{z}_{1}^{3}\overline{z}%
_{2}^{2}.
\end{gather*}

The monomial set:
\begin{alignat*}{3}
& u_{1}=\zeta _{1}^{3}\zeta _{2}^{2},\qquad && v_{1}=z_{1}^{3}z_{2}^{2}, & \\
& u_{2}=\zeta _{1}^{2}\zeta _{2}\zeta _{3},\qquad && v_{2}=z_{1}^{2}z_{2}z_{3}, & \\
& u_{3}=\zeta _{3}\zeta _{4},\qquad && v_{3}=z_{1}^{2}z_{2}z_{3}, & \\
& u_{4}=\zeta _{1}\zeta _{3}^{2},\qquad && v_{4}=z_{1}^{2}z_{2}z_{3}, & \\
& u_{5}=\zeta _{1}\zeta _{2}\zeta _{4},\qquad && v_{5}=z_{1}^{2}z_{2}z_{3}. &
\end{alignat*}

The Hermitian form $\mathbf{L}_{( -1,0) }^{( 1,1)
}( u,\overline{v}) $:
\begin{gather*}
\mathbf{L}_{( -1,0) }^{( 1,1) }( u,\overline{v}%
) =\left(
\begin{array}{c}
v_{1} \\
v_{2} \\
v_{3} \\
v_{4} \\
v_{5}
\end{array}
\right) ^{\dag}\left(
\begin{array}{ccccc}
\frac{7}{144} & \frac{1}{18} & \frac{1}{3} & -\frac{1}{4} & \frac{-1}{12} \vspace{1mm}\\
\frac{1}{18} & \frac{1}{12} & \frac{1}{2} & -\frac{1}{6} & \frac{1}{12} \vspace{1mm}\\
\frac{1}{3} & \frac{1}{2} & 3 & -1 & \frac{1}{2} \vspace{1mm}\\
-\frac{1}{4} & -\frac{1}{6} & -1 & 2 & \frac{3}{2} \vspace{1mm}\\
\frac{-1}{12} & \frac{1}{12} & \frac{1}{2} & \frac{3}{2} & \frac{7}{4}
\end{array}
\right) \left(
\begin{array}{c}
u_{1} \\
u_{2} \\
u_{3} \\
u_{4} \\
u_{5}
\end{array}
\right).
\end{gather*}

Computation of the multiplicity:
\begin{gather*}
\gamma ^{( 1,1) }( ( -1,0) ) =\mathrm{rank}%
\big( \mathbf{L}_{( -1,0) }^{( 1,1) }( u,%
\overline{v}) \big) =2.
\end{gather*}

\subsection*{Acknowledgements}
I would like to express my sincere gratitude to R.~Pnini for his kind help and support.

\pdfbookmark[1]{References}{ref}
\LastPageEnding


\begin{thebibliography}{99}

\footnotesize\itemsep=0pt

\bibitem{ARN-1}
Arnal D., Ben Ammar M., Selmi M.,
Le probl\`{e}me de la r\'{e}duction \`{a} un sous-groupe dans la quantif\/ication par d\'{e}formation,
{\it Ann. Fac. Sci. Toulouse Math. (5)} {\bf 12} (1991), 7--27.

\bibitem{BBCV-1}
Baldoni M.W., Beck M., Cochet C., Vergne M.,
Volume computation for polytopes and partition functions of classical root systems,
{\it Discrete Comput. Geom.} {\bf 35} (2006), 551--595,
\href{http://arxiv.org/abs/math.CO/0504231}{math.CO/0504231}.

\bibitem{Bando-1}
Bando  M., Kuratomo T., Maskawa T., Uehera S.,
Structure of nonlinear realization in supersymmetric theories,
{\it Phys. Lett. B} {\bf 138} (1984), 94--98.

\bibitem{Bando-2}
Bando  M., Kuratomo T., Maskawa T., Uehera S.,
Nonlinear realizations in supersymmetric theories,
{\it Progr. Theoret. Phys.} {\bf 72} (1984), 313--349.\\
Bando  M., Kuratomo T., Maskawa T., Uehera S.,
Nonlinear realizations in supersymmetric theories.~II,
{\it Progr. Theoret. Phys.} {\bf 72} (1984), 1207--1213.

\bibitem{Marinov-1}
Bar-Moshe D., Marinov M.S.,
Realization of compact Lie algebras in K\"ahler manifolds,
{\it J. Phys. A: Math. Gen.} {\bf  27} (1994), 6287--6298,
\href{http://arxiv.org/abs/hep-th/9407092}{hep-th/9407092}.

\bibitem{Marinov-2}
Bar-Moshe D., Marinov M.S.,
Berezin quantization and unitary representations of Lie groups,
in Topics in Statistical and Theoretical Physics, {\it Amer. Math. Soc. Transl. Ser. 2}, Vol.~177,
Amer. Math. Soc., Providence, RI, 1996, 1--21,
\href{http://arxiv.org/abs/hep-th/9407093}{hep-th/9407093}.

\bibitem{Berezin-1}
Berezin F.A.,
Quantization,
{\it Izv. Akad. Nauk SSSR Ser. Mat.} {\bf  38} (1974), 1116--1175.

\bibitem{Berezin-2}
 Berezin F.A.,
 Covariant and contravariant symbols of operators,
{\it Izv. Akad. Nauk SSSR Ser. Mat.} {\bf 36} (1972), 1134--1167.

\bibitem{BGR-1}
Billey S., Guillemin V., Rassart E.,
A vector partition function for the multiplicities of $\mathfrak{sl}_{k}(\mathbb{C})$,
{\it J. Algebra} {\bf 278} (2004), 251--293,
\href{http://arxiv.org/abs/math.CO/0307227}{math.CO/0307227}.

\bibitem{BOU-1}
Baoua O.B.,
Gelfand pairs with coherent states,
{\it Lett. Math. Phys.} {\bf 46} (1998), 247--263.

\bibitem{CAH-1}
Cahen B.,
Multiplicities of compact Lie group representations via Berezin quantization,
{\it Mat.\ Vesnik} {\bf 60} (2008), 295--309.



\bibitem{GS-1}
Guillemin V., Sternberg S.,
Geometric quantization and multiplicities of group representations,
{\it Invent. Math.} {\bf 67} (1982), 515--538.

\bibitem{H-1}
Heckman G.J., Projections of orbits and asymptotic behavior of multiplicities for compact connected Lie groups,
{\it Invent. Math.} {\bf 67} (1983), 333--356.

\bibitem{Hurt-1}
Hurt  N.E.,
Geometric quantization in action,
{\it Mathematics and Its Applications (East European Series)}, Vol.~8, D.~Reidel Publishing Co., Dordrecht-Boston, Mass., 1983.

\bibitem{Kostant-1}
Kostant B.,
A formula for the multiplicity of a weight,
{\it Trans. Amer. Math. Soc.} {\bf 93} (1959), 53--73.

\bibitem{Littelmann-1}
Littelmann P.,
Paths and root operators in representation theory,
{\it Ann of Math. (2)} {\bf 142} (1995), 499--525.

\bibitem{O'Raifeartaigh-1}
O'Raifeartaigh L.,
Group structures of gauge theories,
{\it Cambridge Monographs on Mathematical Physics}, Cambridge University Press, Cambridge, 1986.

\bibitem{Peetre-1}
Peetre J.,
The Berezin transform and Ha-plitz operators,
{\it J. Operator Theory} {\bf 24} (1990), 165--168.

\bibitem{Rawnsley-1}
Rawnsley J., Cahen  M., Gutt~S.,
Quantization of K\"{a}hler manifolds. I.~Geometric interpretation of Berezin's quantization,
{\it J. Geom. Phys.} {\bf 7} (1990), 45--62.

\bibitem{V-1}
Vergne M.,
Multiplicities formula for geometric quantization.~I,
{\it Duke Math. J.} {\bf 82} (1996), 143--179.

\bibitem{V-2}
Vergne M.,
Multiplicities formula for geometric quantization.~II,
{\it Duke Math. J.} {\bf 82} (1996), 181--194.

\end{thebibliography}
\end{document}